# Title: Scrutinizing Data from Sky: An Examination of Its Veracity in Area Based Traffic Contexts

Authors: Yawar Ali (1), Krishnan K N (1), Debashis Ray Sarkar (1), K. Ramachandra Rao (1), Niladri Chatterjee (1), Ashish Bhaskar (2)

((1) Indian Institute of Technology Delhi, New Delhi, India (2) Queensland University of Technology, Brisbane, Australia)

Traffic data collection has been an overwhelming task for researchers as well as authorities over the years. With the advancement in technology and introduction of various tools for processing and extracting traffic data the task has been made significantly convenient. Data from Sky (DFS) is one such tool, based on image processing and artificial intelligence (AI), that provides output for macroscopic as well as microscopic variables of the traffic streams. The company claims to provide 98% to 100 % accuracy on the data exported using DFS tool. The tool is widely used in developed countries where the traffic is homogenous and has lane-based movements. In this study, authors have checked the veracity of DFS tool in heterogenous and area-based traffic movement that is prevailing in most developing countries. The validation is done using various methods using Classified Volume Count (CVC), Space Mean Speeds (SMS) of individual vehicle classes and microscopic trajectory of probe vehicle to verify DFS's claim. The error for CVCs for each vehicle class present in the traffic stream is estimated. Mean Absolute Percentage Error (MAPE) values are calculated for average speeds of each vehicle class between manually and DFS extracted space mean speeds (SMSs), and the microscopic trajectories are validated using a GPS based tracker put on probe vehicles. The results are fairly accurate in the case of data taken from a bird's eye view with least errors. The other configurations of data collection have some significant errors, that are majorly caused by the varied traffic composition, the view of camera angle, and the direction of traffic.

# 1. INTRODUCTION

Traffic data collection is a cornerstone of the transportation studies, providing essential insights for traffic management, planning, and safety analysis. Historically, gathering accurate and detailed traffic data has been a labour-intensive and complex task, loaded with challenges related to accuracy, cost, and scalability. However, the landscape of traffic data collection has undergone a transformative shift with the advent of advanced technologies. It is imperative to have an overview of the utilization of traffic video data in transportation research, along with computational methods and applications relevant to our study. Aerial images are capable of encompassing wide geographical expanses in a single snapshot, making them exceptionally useful for a variety of applications, notably in surveillance and monitoring tasks. This capability has particularly revolutionized methods in observing and analysing traffic patterns, environmental changes, and infrastructure development, among other areas. As a result, there has been a focused shift towards employing unmanned aerial vehicles (UAVs) that can navigate at relatively low altitudes. This development offers a nuanced perspective and greater detail in data collection, facilitating more accurate and comprehensive analyses.

Over the last decade, this area of research has expanded, exploring the potentials of UAV technology in traffic inspection, which includes monitoring traffic flow, detecting congestion, and assessing road conditions, thereby contributing significantly to the advancement of transportation research and planning. In one of the study authors describe detailed exploration of the cutting-edge advancement in the domain of image processing analysis, with a keen emphasis on the utilization of UAVs (Lee & Kwak, 2014).

In recent years, the emergence of image processing and artificial intelligence (AI) has revolutionized the way traffic data is collected, processed, and analyzed. One of the noteworthy advancements in this domain is the development of Data from Sky (DFS), an innovative tool that leverages cutting-edge AI algorithms with their high-end image processing algorithms to extract both macroscopic and microscopic traffic variables from aerial imagery. The tool's ability to provide a comprehensive overview of traffic flow, coupled with its claimed high accuracy rates of 98 to 100 percent, has positioned it as an asset in the area of traffic studies. While DFS has been extensively adopted in countries with homogeneous, lane-based traffic patterns, its effectiveness in heterogeneous, area-based traffic scenarios with prominent seepage behavior (Singh & Ramachandra Rao, 2023), typical of most developing nations, remains an area ripe for exploration. Traffic conditions in most developing countries are characterized by their diversity, with a mix of vehicle types and mostly area-based traffic flow, presenting unique challenges for data collection and analysis. The typical challenges may include the occlusions and shadowing over smaller sized vehicles behind (or beside) large vehicles such as Bus and Heavy Commercial Vehicles (HCVs).

This study aims to address the veracity of working of DFS in area-based heterogenous traffic by rigorously evaluating the performance of the DFS tool in the context of such distinctive traffic conditions. Through a comparative analysis involving Classified Volume Count (CVC), Space Mean Speeds (SMS) for different vehicle categories, and the microscopic trajectory data of probe vehicles, this research scrutinizes the accuracy and reliability of DFS in capturing the nuances of heterogenous traffic composition and area-based traffic dynamics. The methodology employed encompasses a multifaceted approach to validation, where the data obtained from DFS is compared with manually extracted data and GPS-based trajectories to ascertain its veracity. By examining the tool's ability to accurately represent the macroscopic variables and microscopic trajectories of individual vehicles, the study provides a comprehensive assessment of DFS's applicability and precision in a highly heterogenous traffic setting with area-based movement. This analysis not only contributes to the evaluation of an important technological tool but also enhances our understanding of traffic behavior in heterogeneous conditions, offering valuable insights for traffic management and policymaking in similar contexts globally.

## 2. LITERATURE REVIEW

Roadside video cameras (fixed CCTVs) have long been a crucial tool for monitoring traffic. The type of information collected, and the methods used to collect it differs significantly based on the objectives (Kuciemba & Swindler, 2016). Researchers often focus on gathering crucial traffic metrics, such as vehicle counts and movement tracking (Zangenehpour et al., 2015). These data are employed not only to analyse traffic flows but also to conduct safety studies concerning interactions at various traffic corridors (Hu et al., 2004; Morris et al., 2012; St-Aubin et al., 2013, 2015). The ability to identify and monitor moving objects has been a task for researchers specially since after the advent of rapid technological advancement (Buch et al., 2011; Tian et al., 2011).

A Significant amount of research has been carried out in the past two decades on the detection and classification of vehicles using video data. Two major methods are employed from the field of image processing to accomplish this task (Buch et al., 2011). The top-down (TD) approach and bottom-up (BU) approach. TD approach involves identifying vehicle geometry based on their motion characteristics and classifying them into predefined categories through set rules. Key techniques for isolating the foreground include frame differencing, background subtraction, Gaussian mixture models (GMM), and graph cuts, with background subtraction being preferred for its effectiveness (Huang et al., 2017). To classify these foreground elements, various machine-learning techniques are utilized, including artificial neural networks (ANNs), support vector machines (SVMs), and nearest neighbour classifiers. These methods typically train on motion features like corners, edge maps, and optical flow (MacKay, 2012). BU approach vary from the TD approach significantly as those leverage specific object features and detect alterations in pixel values, subsequently categorizing these changes as parts of an object. These components are then aggregated into complete objects for vehicle detection. Techniques for identifying interest points, which are pixel locations where local features are gathered, include the basic path method, scale-invariant feature transform (SIFT), and histogram of oriented gradients (HOG). These interest points are subsequently classified and assimilated into objects using machine learning techniques such as artificial neural networks (ANNs), support vector machines (SVMs), and boosting methods. TD methods have some limitations when it comes to urban traffic due to challenges like misclassification because of the shadows and under (or over) classification of vehicles due to occlusion caused by larger vehicles to smaller following vehicles. BU algorithms are increasingly preferred in heterogenous traffic with area-based movements because they are more adept at managing issues like shadows and occlusions (Li et al., 2013).

Recent advancements have seen cloud computing being employed to construct systems that manage and analyse traffic monitoring data (Abdullah et al., 2014). Systems capable of autonomously processing and analysing video streams through a GPU cluster have been proposed. These systems utilize a cascade classifier for vehicle detection (Abdullah et al., 2014). The techniques employed for vehicle tracking include Kalman filtering, particle filtering, spatial-temporal Markov random fields, and graph correspondence (Coifman et al., 1998; Morse et al., 2016). Analysing driver behaviour from video data entails converting the microscopic trajectories of individual vehicles into comprehensive descriptions of their behaviour and interactions with other vehicles (leader-follower pairs) that primarily relies on detecting vehicles and tracking their movements (Morris & Trivedi, 2013).

The rapid evolution of neural network-based image recognition methods in recent years has led to notable enhancements in detection accuracy (Krizhevsky et al., 2012; Simonyan & Zisserman, 2015). These methods are adept at identifying various types of objects across different scales. Traditional techniques segmented an input image into a fixed grid, where each segment was analysed for a single object label and subsequently merged with adjacent segments (Ciregan et al., 2012; Szegedy et al., 2013). However, contemporary methods leverage neural networks not only to detect objects but also to predict their precise locations, resulting in more accurate and comprehensive outcomes (Redmon et al., 2016; Ren et al., 2015). Deep learning-based methods have recently been introduced for video analytics within smart city applications (Wang & Sng, 2015). One of the libraries gaining significant popularity for image detection is YOLO (Redmon et al., 2016). This library stands out for its capability to detect objects quickly, making it suitable for real-time traffic operations and safety applications. YOLO integrates object detection and recognition into a single neural network model, allowing the algorithm to effectively consider the overall information of a frame while being less affected by shadowing effects (Huang et al., 2017).

While the image processing techniques are developed irrespective of the type of video data, bird's eye view data taken through UAVs is proven to be highly favourable in terms of the output generated. This section sheds light on how drones have transformed the approach to collecting and analysing visual data. It highlights the versatility and efficiency of UAVs in various applications, especially in the field of transportation research. The focus on drones underscores a significant shift towards adopting more advanced, aerial-based methods for image analysis, reflecting a broader trend in leveraging technology to enhance data accuracy and operational efficiency in multiple sectors. Several researchers have compiled a summary of the most recent global research trends concerning the utilization of unmanned aerial vehicles (UAVs), commonly known as drones, for traffic monitoring and analysis (Kanistras et al., 2015; Renard et al., 2022; Salvo et al., 2014; Valavanis & Vachtsevanos, 2015). These studies can be divided into two categories based on two different criteria: (i) equipment type, and (ii) video processing method. Initially, most traffic-related applications involved either mounted video cameras or fixed-wing quadcopter, whereas in recent years only few researchers have initiated investigations using simple multi-rotor UAVs (quadcopter). In the beginning, most traffic-related applications involved fixed-wing UAVs (Barmpounakis et al., 2016; Khan et al., 2017; Renard et al., 2022). Similarly, two main categories of studies have been developed based on the video-processing approach: (i) Semi-Automatic or manual methods, and (ii) Automatic methods. Semi-automatic methods involve precise but laborious and time-consuming processes, as each study requires manual object detection and tracking for many frames (Salvo et al., 2014). However, recently there has been a significant shift towards using automatic methods in transportation studies, resulting in rapid data processing and evolution. This shift has culminated in real-time traffic data analysis captured by UAVs, marking a notable advancement in traffic monitoring and management technologies (Apeltauer et al., 2015; Khan et al., 2017; Renard et al., 2022; Zheng et al., 2015).

The remaining paper is organized in the sequence of methodology, data collection, data processing, data extraction, results, conclusions and references.

## 3. METHODOLOGY

The increasing use of DFS globally has raised questions in the minds of researchers about the validity of DFS output specifically for heterogenous and area-based traffic. The design of experiment for this study is carried out in a way that it can cover variety of datasets for comparison of results. The methodology involves validation of macroscopic as well as microscopic variables in the traffic stream. That approach resulted in a more robust outcome and generalized conclusions for traffic prevailing in most developing nations.

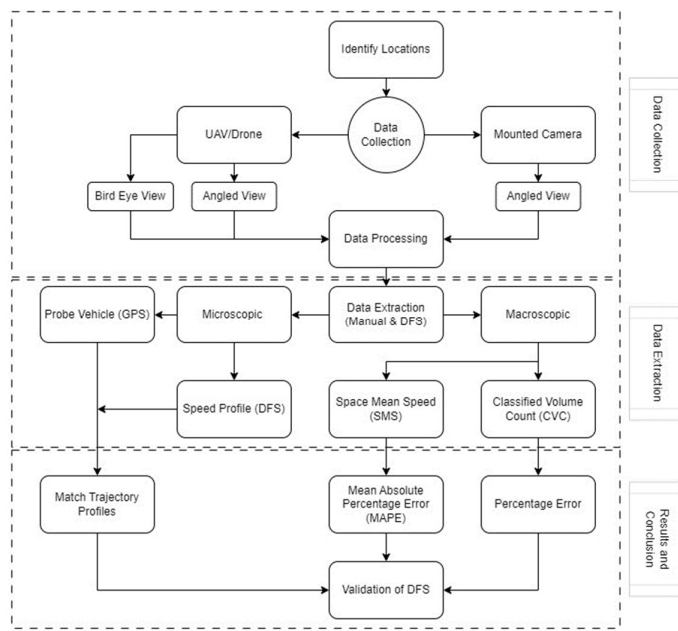

*Figure 1 Methodology*

Firstly, four locations are identified where data could be collected using UAV/mounted camera. Four locations leading to 8 data sets (1 location - 2 directions of traffic) are identified. The SMS values are used to calculate Mean Absolute Percentage Error (MAPE) for a sample of 100 vehicles from each class. Lastly, the microscopic speed profiles are matched for the probe vehicle from GPS output and DFS output. The results are reported, and conclusions are made in the final section of this study.

## 4. DATA COLLECTION

The data collection is carried out through the design of experiments (DoE) for this study. As per the DoE it is finalized to have four independent data sets of traffic videos. The specifications of the data sets with the desired variability are shown in Table 1. The data is collected using two distinct tools. One being the Unmanned Aerial Vehicle (UAV) or Drone namely DJI Mini 2 which is capable of shooting 4K resolution videos at 30 frames per second (fps). The other being GoPro Hero 7 which is capable of shooting 4K resolution videos at 25 fps. One of the drawbacks of using UAVs for data collection is that the battery backup ranges between 15 to 20 minutes which results in lag in the continuous dataset if it is of higher duration. Four locations are identified for data collection in the urban region of Delhi-NCR in North India.

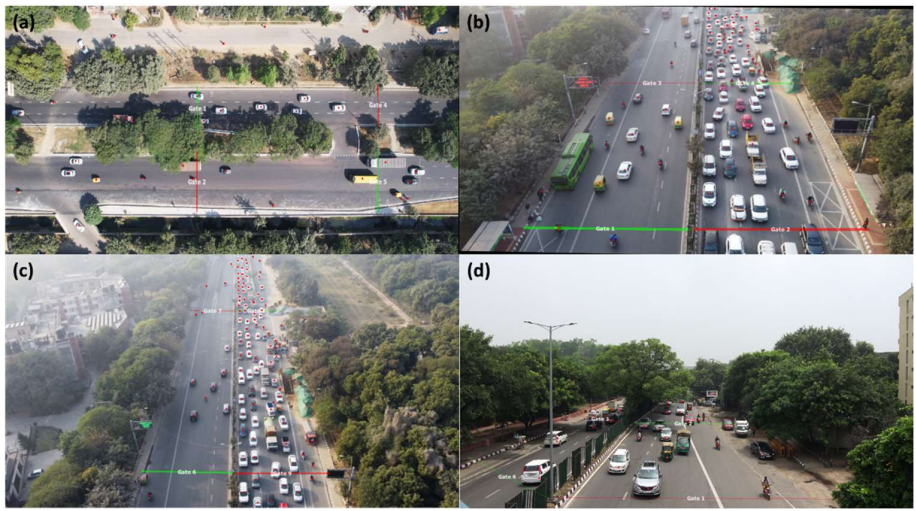

*Figure 2 (a) Bird's Eye Data (b) Angled Data at 25m Height (c) Angled Data at 50m Height (d) Mounted Camera Data*

The data sets are collected at urban arterials at various locations in the national capital region (NCR) of India. The first data (Figure 2: a) is collected using a drone at 150 m height from the road surface with a bird's-eye view. The second and third (Figure 2: b and c) data is collected using a drone at 25 m and 50 m height respectively from the ground at an angled view of 60° from the horizontal. The last dataset (Figure 2: d) is collected using a mounted camera pinned at a foot over bridge (FoB) height (5.5 m) with the view angle of about 30 degrees from the horizontal. Each dataset is collected for a duration of 30 minutes.

The specifications of the collected data can be seen in Table 1. The bird's eye view data has two directions of traffic, viz., East Bound and West Bound, rest three of the data sets have traffics in North Bound and South Bound directions. For consistency in reporting the results, the first data set East bound (Top half of the frame) will be referred to as North Bound (NB) and the West Bound (Bottom half of the frame) will be referred to as South Bound (SB).

*Table 1 Specifications of Data Collection*

| TOOL(S) | View Angle (Degrees) | Height (Meters) | Duration (~Minutes) | Extraction Tool |
|---|---|---|---|---|
| UAV | 90 (Bird's-Eye) | 150 | 30 | DFS – Aerial |
| UAV | 60 | 50 | 30 | DFS – Aerial |
| UAV | 60 | 25 | 30 | DFS – Aerial |
| Mounted Camera | 30 | 5.5 | 30 | DFS – Light |

## 5. DATA PROCESSING

Data collected using UAV in its raw form is not feasible to work with, as the drone clips the recordings in approximately 5.5 minutes duration to avoid data loss in case of drone malfunctioning. The data collected using UAV needs to be stitched to make a single video for post processing. The video clips are stitched together using iMovies software for further processing. The 4 sets of videos are uploaded to be processed through DFS on their website and simultaneously the videos are used for manual extraction of the required data.

## 6. DATA EXTRACTION

### 5.1 Manual Extraction

The manual extraction of CVC is done by marking a line on the road section as a gate. Every time the vehicle crosses the gate it is recorded along with its classification. The cumulative count of each vehicle class is recorded for all the data sets for a duration of 30 minutes each. The classification of vehicles reported in this study are Motorized Two Wheelers (M2W), Motorized Three Wheelers (M3W), Cars, Light Commercial Vehicles (LCV), Bus, and Heavy Commercial Vehicles (HCV).

The SMS calculations are done by marking the two gates (Figure 2) on the video and making a trap length already known on the field. Randomly identified vehicles of different classes are recorded their entry and exit time stamps in the marked trap length. The travel time of each vehicle is calculated by subtracting the exit and entry time stamp. The trap length for each location is fixed (Figure 2: (a) 62m, (b) 100m, (c) 47m, (d) 32m), so the SMS is calculated by dividing the trap length by the travel time. The variation in trap lengths also takes into account the sensitivity of SMS calculations for DFS output. The same is done for a sample of 100 vehicles from each class. Some vehicle classes, like Bus, had less population in the data so the population is used in the SMS calculations.

### 5.2 Extraction with Data from Sky

Data extraction with DFS is not very straightforward as it requires some steps of pre-processing to be followed. Firstly, the stitched video needs to be uploaded to DFS website for processing (paid). After the video is uploaded it takes a few hours to get processed and the processed output file must be downloaded that runs on DFS viewer only, which is a proprietary software.

The CVC can be extracted for all the vehicles in frame by simply clicking the "Export Traffic Analysis to Excel" to get the statistics in an excel sheet. Since we are not interested in all the vehicles present in the frame through the video duration, we need to mark a gate in the DFS viewer and extract the "gate crossing events" in a spreadsheet. The details of the steps involved in operating the DFS viewer are not described here, as the step wise instructions are found in the manual (https://intercom.help/datafromsky/en/collections/1997337-trafficsurvey-post-recording-processing-of-videos).

The extraction of speeds requires a set of specific steps. Firstly, the video needs to be calibrated in the DFS viewer using the real-world coordinates. DFS uses the term Geo-registration for this step. Doing this step makes the software understand the reference points with respect to which the speeds are estimated in the recorded video. Secondly, the annotation configurations of the file must be modified by adding entry and exit gates on the video frame as per the known length of stretch on the field. Once the video is calibrated and annotated with the least error, we can extract the SMS of the required vehicle IDs (same as the Manual Extraction) from the DFS viewer by exporting the gate crossing trajectories.

## 7. RESULTS

### 6.1 Macroscopic

The macroscopic variables that are chosen to be used for validation of DFS in heterogenous traffic are Classified Volume count (CVC) and Space Mean Speeds (SMS).

### 7.1.1 Classified Volume Count (CVC)

The CVC output for manual extraction and DFS show promising results for the majority proportion of traffic stream i.e., M2W, M3W and Cars which combined is over 90% of the fleet. Figure 3 shows the class wise variation in CVC using two methods of extraction, viz., Manual and DFS. The spider plot vertices represent the datasets, which is 8 in our study and the contours of the plot represent the cumulative count of CVC. The manual and DFS output overlapping in the plot represent the least variation of CVC by two extraction methods.

Interestingly, there is a high discrepancy between the outputs for HCV and Buses class. That too is not seen for all the 8 datasets. To understand the discrepancy in these values, the percentage error ($e_p$) is estimated for each vehicle class and each of the eight datasets using equation (1).

$$e_p = \frac{(Manual\ CVC - DFS\ CVC)}{(Manual\ CVC)} * 100 \quad (1)$$

The estimated error values are tabulated and are shown in Table 2. The nomenclature for the data sets is as, Bird's Eye view data (90°), Angled Data (50m-60° and 25m-60°) which represent the height of UAV and camera view angle, and lastly Mounted camera data set.

The directions of traffic for each data set are reported as North Bound (NB) or South Bound (SB) as defined earlier in Data Collection section of this paper. The two methods of extraction are reported as Manual and DFS. The negative values of $e_p$ in the table indicate the percentage of false positives in the data while the positive values indicate percentage of false negatives in the datasets.

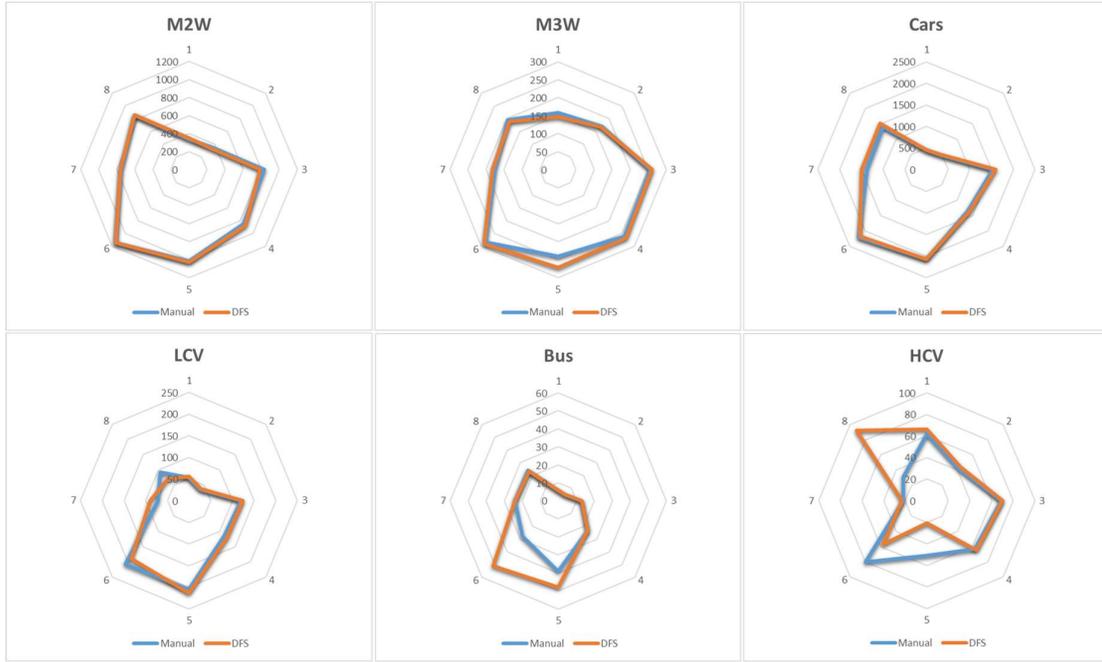

*Figure 3 Visualisation of the CVC difference between Manual and DFS output*

To discuss each data set individually, we observe that Bird's Eye view data set and Mounted camera data set are showing promising results with low classification error. When it comes to the angled data at 50 m height and 60-degree camera view angle, there are relatively high classification errors especially in the case of Buses and HCVs.

*Table 2 Percentage Error in CVC*

| Vehicle | Manual 90° | | DFS 90° | | $e_p$ (%) | | 50m-60° Manual | | 50m-60° DFS | | $e_p$ (%) | |
|---|---|---|---|---|---|---|---|---|---|---|---|---|
| | NB | SB | NB | SB | NB | SB | NB | SB | NB | SB | NB | SB |
| M2W | 336 | 356 | 349 | 336 | -3.9 | 5.6 | 1012 | 1150 | 1026 | 1148 | -1.4 | 0.2 |
| M3W | 158 | 171 | 147 | 167 | 7.0 | 2.3 | 240 | 285 | 272 | 290 | -13.3 | -1.8 |
| Cars | 461 | 473 | 464 | 485 | -0.7 | -2.5 | 2066 | 2216 | 2074 | 2170 | -0.4 | 2.1 |
| LCV | 53 | 36 | 56 | 38 | -5.7 | -5.6 | 205 | 207 | 213 | 189 | -3.9 | 8.7 |
| Bus | 6 | 5 | 6 | 5 | 0.0 | 0.0 | 39 | 28 | 48 | 51 | -23.1 | -82.1 |
| HCV | 62 | 42 | 66 | 44 | -6.5 | -4.8 | 51 | 80 | 21 | 57 | 58.8 | 28.8 |
| | Mounted Manual | | Mounted DFS | | $e_p$ (%) | | 25m-60° Manual | | 25m-60° DFS | | $e_p$ (%) | |
| | NB | SB | NB | SB | NB | SB | NB | SB | NB | SB | NB | SB |
| M2W | 836 | 855 | 790 | 881 | 5.5 | -3.0 | 751 | 852 | 763 | 854 | -1.6 | -0.2 |
| M3W | 256 | 261 | 261 | 267 | -2.0 | -2.3 | 175 | 196 | 182 | 190 | -4.0 | 3.1 |
| Cars | 1530 | 1343 | 1598 | 1391 | -4.4 | -3.6 | 1370 | 1409 | 1500 | 1514 | -9.5 | -7.5 |
| LCV | 119 | 115 | 124 | 122 | -4.2 | -6.1 | 72 | 93 | 90 | 70 | -25.0 | 24.7 |
| Bus | 13 | 23 | 13 | 23 | 0.0 | 0.0 | 24 | 24 | 24 | 23 | 0.0 | 4.2 |
| HCV | 69 | 63 | 70 | 65 | -1.4 | -3.2 | 22 | 31 | 23 | 92 | -4.5 | -197 |

The NB direction error for M3W and Bus reflects that the classification is giving false positives. On manual inspection it is realized that the vehicle moving away from the camera where an M3W is sometimes classified as LCV, and where a Bus is classified as HCV, and interchangeably so. For the SB direction, when the vehicle is moving towards the camera, the same type of classification error is observed between Bus and HCV. Since the traffic has varying structure of HCVs, some may appear as Bus if looked at from an angled view. Similar kind of patterns in classification error are observed for the data with reduced height of 25 m and 60-degree view angle. Since in this case, the traffic is viewed better as the coverage area of UAV is reduced, the error for Bus is resolved but the LCV errors increased, and HCV error shot up to almost 200%. On manual inspection, it is realized that a significant share of LCV is wrongly classified as HCV in the congested traffic stream (NB), and since the count values are low, the error is magnified. Even in this case, the NB direction having a free flow traffic shows less error for HCV despite the vehicle movement away from the camera. This leads to the conclusion that the state of traffic stream also plays an important role along with the direction of movement of traffic.

### 7.1.2 Space Mean Speed (SMS)

Space Mean Speeds (SMS) are a good macroscopic measure to use for validation in this study since we are using space sensors to collect the data. Mean Absolute Percentage Error (MAPE) values are used to validate the SMS of DFS output with the manually calculated ones.

$$MAPE\ (\%) = 100 * \frac{1}{n} \sum_{i=1}^{n} \frac{|v_s^{DFS} - v_s^m|}{v_s^m} \qquad (2)$$

Where, n is Number of samples, i is Vehicle Class, $v_s^{DFS}$ is SMS from DFS output, and $v_s^m$ is Manually Calculated SMS.

MAPE values under 10% are generally considered good for any experimental design. In our study, all the MAPE values lie well under 10% (Table 3) that suggests that the overall accuracy of DFS is good. Since all the values are well within the desired limit, we will be discussing the results in relative terms. Starting with the mounted camera data, we see that the MAPE values are relatively higher on the NB direction of traffic. This difference is related to the fact that the view of NB direction of traffic is skewed as compared to the SB direction of traffic.

*Table 3 MAPE values for SMS*

| Data | Mounted | | 90° | | 25m-60° | | 50m-60° | |
|---|---|---|---|---|---|---|---|---|
| | NB | SB | NB | SB | NB | SB | NB | SB |
| Vehicle | MAPE (%) | | | | | | | |
| M2W | 7.11 | 4.4 | 1.92 | 3.49 | 2.25 | 2.85 | 1.91 | 5.27 |
| M3W | 7.07 | 4.82 | 3.63 | 1.97 | 2.1 | 2.7 | 2.12 | 4.8 |
| Cars | 5.72 | 3 | 2.61 | 1.8 | 2.22 | 2.04 | 2.17 | 5.95 |
| LCV | 7.13 | 5.4 | 1.8 | 1.3 | 2.24 | 2.57 | 2.13 | 5.92 |
| Bus | 4.8 | 6.85 | 1.65 | 1.65 | 1.58 | 2.54 | 2.52 | 4.92 |
| HCV | 6.9 | 5.02 | 1.95 | 1.95 | 2.03 | 2.9 | 1.97 | 5.18 |

The bird's eye view data gives the best results in terms of the MAPE values which are least amongst all four of data locations. This is likely because calibrating the video with real world coordinates in bird's eye data is less complex than other views, as it requires least correction for the frame perspective. The angled data taken at 25 m height gives better values of MAPE as compared with the angled data taken at 50 m height. This observation leads to the inference that a lower height angled data that covers lesser stretch of road gives more accurate results, hence, there is a tradeoff between road stretch coverage and accuracy.

An interesting thing to note in the angled data from 50 m height is that the SB direction has higher MAPE values than the NB direction due to the heavy volume (congested traffic) of traffic on the SB direction. The effect on MAPE of traffic state is not apparent in the other angled data from 25 m height. In SMS estimation the combined effect of angle of view and traffic state is more on the data taken from greater height.

### 6.2 Microscopic

Data from Sky has various features to work with and microscopic vehicle trajectories is one of them. Since the application of microscopic vehicle trajectories is quite essential for various traffic related studies, especially for safety analyses, it is imperative to validate the trajectories exported through the DFS software.

**Probe Vehicle Trajectory**

Validating the trajectories of DFS requires a simultaneous process of data collection to have microscopic trajectories from two different methods. One method being the reference method is chosen to be Global Positioning System (GPS) based trajectories. An android based application named Geo Tracker is used for taking ten unique samples of probe vehicle trajectories. While the GPS trajectories were being recorded, a drone was set up at a height of 60 m with a bird's eye view to capture the same vehicle's movement.

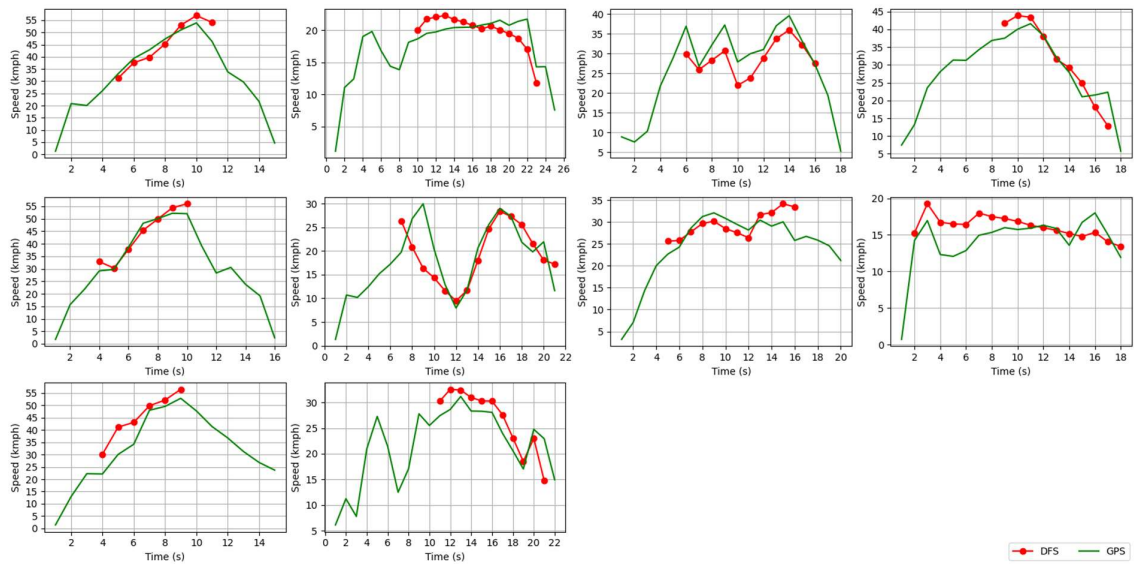

*Figure 4 Microscopic Trajectory Validation*

The GPS based trajectories were recorded in a Keyhole Markup Language (KML) file and the data was extracted from the KML file using a python script. Simultaneously, the drone recorded video was processed through DFS website and the microscopic trajectories were extracted from DFS viewer. The trajectories from DFS viewer need to be extracted after doing geo-registration. Out of the trajectories of all the vehicles, the probe vehicle ID is identified in the stream and extracted for comparison with the GPS trajectories of the same vehicle.

Once the trajectories are extracted, the speed profile of the probe vehicle is plotted against time (Figure 4) for both GPS and DFS output. Microscopic validation can be done subjectively or objectively (Benekohal, 1991). Pertaining to subjective validation, the graphical comparison of speed profiles can be a good indicator. The resulting graphs were promising in terms of their match of the microscopic trajectories. All the ten unique samples showed only marginal errors in the speed profiles, as can be seen in the graphical plots. Concerning the objective validation (using statistical techniques) of microscopic trajectories, a parametric test was performed pertaining to the differences between paired observations being normally distributed. A paired t-test was performed with the null hypothesis being 'equal means' of speeds attained from DFS and GPS. The level of significance for the test was kept being 0.05 ($\alpha$). The Pearson correlation value came out to be 0.94, suggesting a very strong positive linear relationship between the two variables. The p-value for two tails is estimated to be 0.42 ($>\alpha$), hence the null hypothesis is accepted. The hypothesis test suggests that there is no significant difference in the speeds attained from DFS and GPS, hence validating the DFS output objectively.

## 8. CONCLUSIONS

The conclusions are made based on the experiment designed for this study. The claim that the DFS has a 98% to 100% accuracy, needs to be qualified under highly heterogenous and area-based traffic conditions. The results have been fairly accurate in the case of data taken with a UAV – bird's eye view (Table 3). Another advantageous feature of using UAV data is that a longer stretch of road can be captured at once with the least error in DFS output. The results of the angled data collected at different heights suggest that the DFS is not well trained yet for the highly heterogenous traffic composition. In the case of angled data, the direction of traffic movement (towards or away from camera) as well as the traffic stream state (free flow or congested) seem to affect the efficacy of DFS software. In terms of the microscopic vehicle trajectories that are validated subjectively as well as objectively, the DFS algorithm seems to work well, provided the calibration of the video on DFS viewer (Geo-Registration) is done precisely with least error.

The findings of this study yield actionable recommendations for researchers in the field of transportation studies, who are utilizing DFS as a tool for data processing, to opt for UAV (bird's eye) data to have higher accuracy in the sought output from DFS in heterogenous traffic composition with area-based movement.

The study has a few limitations, like the veracity of using DFS for nighttime, adverse wind and foggy weather condition data is not captured. It is limited to the data collected at midblock locations only. Intersections and roundabouts can be included in future scope.

## 9. AUTHORS CONTRIBUTION STATEMENT

All the authors have conceptualized and designed the study. The first, second & third authors participated in data collection. The first author processed and analysed the data and prepared the results. Each author has carried out interpretation of the results. The first author prepared the draft manuscript. All authors reviewed the results and approved the final version of the manuscript.

## 10. CONFLICT OF INTEREST

The authors certify that they have NO affiliation with or involvement in any organization or entity with any financial interest or non-financial interest in the subject matter or materials discussed that could have appeared to influence the work reported in this manuscript.

## 11. DATA AVAILABILITY STATEMENT

The data that support the findings of this study are available on reasonable request from the corresponding author.

## 12. REFERENCES

Abdullah, T., Anjum, A., Tariq, M. F., Baltaci, Y., & Antonopoulos, N. (2014). Traffic monitoring using video analytics in clouds. *Proceedings - 2014 IEEE/ACM 7th International Conference on Utility and Cloud Computing, UCC 2014*. https://doi.org/10.1109/UCC.2014.12

Apeltauer, J., Babinec, A., Herman, D., & Apeltauer, T. (2015). Automatic vehicle trajectory extraction for traffic analysis from aerial video data. *International Archives of the Photogrammetry, Remote Sensing and Spatial Information Sciences - ISPRS Archives*, *40*(3W2), 9–15. https://doi.org/10.5194/isprsarchives-XL-3-W2-9-2015

Barmpounakis, E. N., Vlahogianni, E. I., & Golias, J. C. (2016). Unmanned Aerial Aircraft Systems for transportation engineering: Current practice and future challenges. *International Journal of Transportation Science and Technology*, *5*(3), 111–122. https://doi.org/10.1016/j.ijtst.2017.02.001

Benekohal, R. F. (1991). Procedure for Validation of Microscopic Traffic Flow Simulation Models. *Transportation Research Record*.

Buch, N., Velastin, S. A., & Orwell, J. (2011). A review of computer vision techniques for the analysis of urban traffic. In *IEEE Transactions on Intelligent Transportation Systems*. https://doi.org/10.1109/TITS.2011.2119372

Ciregan, D., Meier, U., & Schmidhuber, J. (2012). Multi-column deep neural networks for image classification. *Proceedings of the IEEE Computer Society Conference on Computer Vision and Pattern Recognition*. https://doi.org/10.1109/CVPR.2012.6248110

Coifman, B., Beymer, D., McLauchlan, P., & Malik, J. (1998). A real-time computer vision system for vehicle tracking and traffic surveillance. *Transportation Research Part C: Emerging Technologies*. https://doi.org/10.1016/S0968-090X(98)00019-9

Hu, W., Xiao, X., Xie, D., Tan, T., & Maybank, S. (2004). Traffic accident prediction using 3-D model-based vehicle tracking. *IEEE Transactions on Vehicular Technology*. https://doi.org/10.1109/TVT.2004.825772

Huang, L., Xu, W., Liu, S., Pandey, V., & Juri, N. R. (2017). Enabling versatile analysis of large scale traffic video data with deep learning and HiveQL. *Proceedings - 2017 IEEE International Conference on Big Data, Big Data 2017*. https://doi.org/10.1109/BigData.2017.8258041

Kanistras, K., Martins, G., Rutherford, M. J., & Valavanis, K. P. (2015). Survey of unmanned aerial vehicles (uavs) for traffic monitoring. *Handbook of Unmanned Aerial Vehicles*, 2643–2666. https://doi.org/10.1007/978-90-481-9707-1_122

Khan, M. A., Ectors, W., Bellemans, T., Janssens, D., & Wets, G. (2017). Unmanned aerial vehicle-based traffic analysis: Methodological framework for automated multivehicle trajectory extraction. *Transportation Research Record*, *2626*, 25–33. https://doi.org/10.3141/2626-04

Krizhevsky, A., Sutskever, I., & Hinton, G. E. (2012). ImageNet classification with deep convolutional neural networks. *Advances in Neural Information Processing Systems*.

Kuciemba, S., & Swindler, K. (2016). *Transportation Management Center Video Recording and Archiving Best General Practices*. https://ops.fhwa.dot.gov/publications/fhwahop16033/fhwahop16033.pdf

Lee, J., & Kwak, K. (2014). A Trends Analysis of Image Processing in Unmanned Aerial Vehicle. *International Journal of Computer, Information, Systems and Control Engineering*, *8*(2), 261–264.

Li, Y., Li, B., Tian, B., & Yao, Q. (2013). Vehicle detection based on the and- or graph for congested traffic conditions. *IEEE Transactions on Intelligent Transportation Systems*. https://doi.org/10.1109/TITS.2013.2250501

MacKay, D. J. C. (2012). Information Theory, Inference, and Learning Algorithms. In *Advanced Science Letters*.

Morris, B. T., Tran, C., Scora, G., Trivedi, M. M., & Barth, M. J. (2012). Real-time video-based traffic measurement and visualization system for energy/emissions. *IEEE Transactions on Intelligent Transportation Systems*. https://doi.org/10.1109/TITS.2012.2208222

Morris, B. T., & Trivedi, M. M. (2013). Understanding vehicular traffic behavior from video: a survey of unsupervised approaches. *Journal of Electronic Imaging*. https://doi.org/10.1117/1.jei.22.4.041113

Morse, P., St-Aubin, P., Miranda-Moreno, L., Saunier, N., & Board, T. R. (2016). *Transferability Study of Video Tracking Optimization for Traffic Data Collection and Analysis* (p. 20p). https://trid.trb.org/view/1394500

Redmon, J., Divvala, S., Girshick, R., & Farhadi, A. (2016). You only look once: Unified, real-time object detection. *Proceedings of the IEEE Computer Society Conference on Computer Vision and Pattern Recognition*. https://doi.org/10.1109/CVPR.2016.91

Ren, S., He, K., Girshick, R., & Sun, J. (2015). Faster R-CNN: Towards real-time object detection with region proposal networks. *Advances in Neural Information Processing Systems*.

Renard, A., Novacko, L., Babojelic, K., & Kozul, N. (2022). Analysis of Child Traffic Safety near Primary School Areas Using UAVTechnology. *SUSTAINABILITY*, *14*(3). https://doi.org/10.3390/su14031144

Salvo, G., Caruso, L., & Scordo, A. (2014). Urban Traffic Analysis through an UAV. *Procedia - Social and Behavioral Sciences*, *111*, 1083–1091. https://doi.org/10.1016/j.sbspro.2014.01.143

Simonyan, K., & Zisserman, A. (2015). Very deep convolutional networks for large-scale image recognition. *3rd International Conference on Learning Representations, ICLR 2015 - Conference Track Proceedings*.

Singh, M. K., & Ramachandra Rao, K. (2023). Simulation of Signalized Intersection with Non-Lane-Based Heterogeneous Traffic Conditions Using Cellular Automata. *Transportation Research Record*. https://doi.org/10.1177/03611981231211317

St-Aubin, P., Miranda-Moreno, L., & Saunier, N. (2013). An automated surrogate safety analysis at protected highway ramps using cross-sectional and before-after video data. *Transportation Research Part C: Emerging Technologies*. https://doi.org/10.1016/j.trc.2013.08.015

St-Aubin, P., Saunier, N., & Miranda-Moreno, L. (2015). Large-scale automated proactive road safety analysis using video data. *Transportation Research Part C: Emerging Technologies*. https://doi.org/10.1016/j.trc.2015.04.007

Szegedy, C., Toshev, A., & Erhan, D. (2013). Deep Neural Networks for object detection. *Advances in Neural Information Processing Systems*. https://doi.org/10.54097/hset.v17i.2576

Tian, B., Yao, Q., Gu, Y., Wang, K., & Li, Y. (2011). Video processing techniques for traffic flow monitoring: A survey. *IEEE Conference on Intelligent Transportation Systems, Proceedings, ITSC*. https://doi.org/10.1109/ITSC.2011.6083125

Valavanis, K. P., & Vachtsevanos, G. J. (2015). Handbook of unmanned aerial vehicles. *Handbook of Unmanned Aerial Vehicles*, 1–3022. https://doi.org/10.1007/978-90-481-9707-1

Wang, L., & Sng, D. (2015). *Deep Learning Algorithms with Applications to Video Analytics for A Smart City: A Survey*. https://arxiv.org/abs/1512.03131v1

Zangenehpour, S., Miranda-Moreno, L. F., & Saunier, N. (2015). Automated classification based on video data at intersections with heavy pedestrian and bicycle traffic: Methodology and application. *Transportation Research Part C: Emerging Technologies*. https://doi.org/10.1016/j.trc.2015.04.003

Zheng, C., Breton, A., Iqbal, W., Sadiq, I., Elsayed, E., & Li, K. (2015). Driving-behavior monitoring using an Unmanned Aircraft System (UAS). *Lecture Notes in Computer Science (Including Subseries Lecture Notes in Artificial Intelligence and Lecture Notes in Bioinformatics)*, *9185*, 305–312. https://doi.org/10.1007/978-3-319-21070-4_31